\documentclass[letterpaper,10pt]{article}
\usepackage[utf8]{inputenc}

\usepackage{color}
\usepackage{xcolor}
\usepackage{graphicx}
\usepackage{tikzscale}
\usepackage{tikz}
\usepackage{pgfplots}
\usetikzlibrary{plotmarks,matrix,chains,scopes,fit,calc,shapes,positioning,decorations,intersections,fit,backgrounds,patterns}
\pgfplotsset{compat=newest} 

\usepgfplotslibrary{external}
\tikzsetexternalprefix{tikz/}
\tikzsetfigurename{tikz} 
\tikzset{external/up to date check=md5}

\pgfplotsset{plot coordinates/math parser=false}
\pgfplotsset{every axis plot/.append style={solid,line width=1.5pt,mark size=1.5pt,mark options={solid,fill=white}}}
\pgfplotsset{every axis legend/.append style={legend cell align=left,font=\footnotesize}}

\colorlet{42GBd16QAM_color}{blue!80!white}
\colorlet{42GBd64QAM_color}{red!80!black}
\colorlet{64GBd16QAM_color}{orange!90!black}
\colorlet{64GBd64QAM_color}{green!50!black}

\pgfplotsset{64GBd64QAM/.style={color=64GBd64QAM_color,solid}}
\pgfplotsset{64GBd16QAM/.style={color=64GBd16QAM_color,dotted}}
\pgfplotsset{42GBd64QAM/.style={color=42GBd64QAM_color,dashed}}
\pgfplotsset{42GBd16QAM/.style={color=42GBd16QAM_color,dash dot dot}}

\newlength\FigureWidth
\newlength\FigureHeight
\newlength\FullFigureWidth
\graphicspath{{figures/}}

\pgfplotsset{myLegend/.append style={legend style={font=\footnotesize,at={(0.5,0.98)},anchor=north,align=left,legend columns=3}}}

\newcommand{\inputtikz}[1] {
  \includegraphics{tikz/#1.pdf}
}

\newcommand{\Rloss}{\ensuremath{R_{\text{loss}}}\xspace}
\newcommand{\Rlossn}{\ensuremath{R_{\text{loss},n}}\xspace}
\newcommand{\AIRn}{\ensuremath{\text{AIR}_n}\xspace}

\usepackage{osameet3} 
\usepackage{amsmath,amssymb}
\usepackage[normalem]{ulem}

\usepackage{xspace}
\usepackage{wrapfig}
\usepackage{caption}

\usepackage[colorlinks=true,bookmarks=false,citecolor=blue,urlcolor=blue]{hyperref} 

\begin{document}

\title{Mitigating Fiber Nonlinearities by\\Short-length Probabilistic Shaping}

\author{Tobias Fehenberger, Helmut Griesser, and Jörg-Peter Elbers}
\address{ADVA, Fraunhoferstr. 9a, 82152 Martinsried/Munich, Germany\\}
\vspace{-3pt}
\email{\href{mailto:tfehenberger@adva.com}{tfehenberger@adva.com}}
\vspace{-14pt}
\begin{abstract} 
We show that short-length probabilistic shaping reduces nonlinear interference in optical fiber transmission. SNR improvements of up to 0.8 dB are obtained. The shaping gain vanishes when interleaving is employed and not undone before transmission.\\
\end{abstract}

\vspace{-8pt}
\section{Introduction}
\vspace{-7pt}

  Probabilistic shaping (PS) is a widely employed technique to achieve signal-to-noise ratio (SNR) gains of more than 1~dB for high-order quadrature amplitude modulation (QAM) and to enable rate adaptivity with fixed QAM order and forward error correction (FEC) overhead. Probabilistic amplitude shaping (PAS) has become the de-facto standard to realize PS \cite{GeorgTComm}. The effect of PAS on nonlinear interference (NLI) in optical fiber transmission has attracted particular attention (see e.g. \cite{myJLT, JulianJLT, SillikensShaping}). Typically, the shaped signaling scheme was emulated by randomly drawing QAM symbols according to the desired shaped distribution, neglecting the influence of the distribution matcher (DM) that maps uniform bits to shaped amplitudes.


  Recent simulations \cite{ESS_Karim} and experiments \cite{ESS_Sebastiaan} found that the effective SNR after fiber transmission and digital signal processing (DSP) increases when the block length of the employed DM scheme is reduced. Such an effect had not been observed before, mainly due to random-draw based PAS emulation technique used previously. The block-length dependence of SNR has been studied in detail in \cite{myCCJLT2019} for variable-length outputs of a constant-composition DM (CCDM) \cite[Sec.~V]{GeorgTComm}. It was found that temporal properties of the QAM transmit sequence that are guaranteed to occur for short CCDM outputs yet not for long blocks lead to NLI mitigation. This effect is related to previous theoretical work on shaping for the nonlinear fiber channel predicting larger gains than for a linear channel \cite{RonenShaping} and also to temporal PS over a few time slots \cite{GellerShaping,MetodiTemporalShaping}.

  In this manuscript, temporal shaping effects are studied that are inherently introduced by short-length PS, extending previous work on the subject \cite{myCCJLT2019}. We observe a mitigation of NLI and thus significant SNR improvements in fiber simulations for shaped and also for uniformly distributed QAM sequences that are generated by a CCDM. These improvements are only present in end-to-end simulations in which the DM is fully implemented, and not when the PAS framework is just emulated. To exploit the shaping gains, correlations within the QAM transmit sequences must be preserved and any interleaving must be avoided or undone before transmission.
  

\vspace{-5pt}
\section{Probabilistic amplitude shaping: End-to-end vs. emulation}
\begin{figure}
\begin{center}
\inputtikz{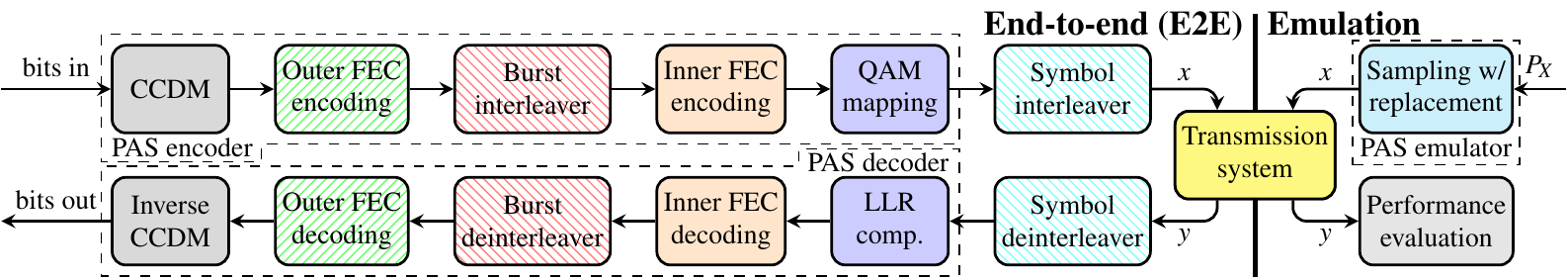}
\captionsetup{width=5.5in}
\vspace*{-1.4\baselineskip}
\caption{Block diagrams of PAS transmission systems. Left: End-to-end (E2E) implementation with PAS encoder and decoder. Two interleavers are included, as proposed for 400ZR \cite{400ZR}. Note that the outer FEC and the interleavers (hatched blocks) are often omitted in E2E implementations. Right: Emulation with QAM symbols directly drawn according to the desired shaped distribution $P_X$.}
\label{fig:block_diagram}
\end{center}
\vspace*{-2.5\baselineskip}
\end{figure}

\vspace{-7pt}
Figure~\ref{fig:block_diagram} shows the key blocks for PS signaling in simulations or equivalently in experiments. Depicted on the left is an end-to-end setup with a full PAS encoder and decoder \cite{ESS_Sebastiaan}. The encoder comprises the amplitude shaping block (in our case a CCDM) that maps uniform data bits to shaped amplitudes, as well as one or more systematic FEC encoders that preserve the information bit sequences at their respective outputs. In the considered case of a concatenated two-stage FEC, a burst interleaver is required between the inner and outer FEC to spread an unsuccessfully decoded inner FEC codeword over several outer codewords. The inner FEC output is mapped to shaped QAM symbols $x$, which are again interleaved to achieve polarization- and phase-diversity as proposed for 400ZR \cite{400ZR}. The sequences are then sent over an optical system comprising the transmitter, the fiber link, and the receiver including DSP. After symbol deinterleaving, log-likelihood ratios (LLRs) are computed in the PAS decoder and used for decoding of the inner code. The output is deinterleaved, decoded, and an inverse CCDM recovers the initially sent data bits. 

A greatly simplified setup for emulating PS signaling is shown on the right of Fig.~\ref{fig:block_diagram}. The entire PAS encoder is emulated by a single block that performs sampling with replacement, i.e., the QAM symbols $x$ are randomly drawn according to the target distribution $P_X$ without including the CCDM or the FEC. The transmit sequence is sent over the same optical system as above. After DSP, performance metrics are evaluated, e.g. by computing SNR or achievable information rates (AIRs). This emulation approach ignores the underlying coded modulation architecture for realizing PS signaling, as it omits several key building blocks. Nonetheless, it has proven useful for studying the impact of shaped QAM signaling on various blocks of an optical communication system, such as the digital-to-analog converter, fiber nonlinearities, or the receiver DSP. Certain nonlinear effects that lead to a block-length dependent SNR, though, are not captured by this emulation, as we will show in the following.

\vspace{-5pt}
\section{Numerical Simulations}\label{sec:sim_setup}
\vspace{-5pt}
\subsection{Setup}
\vspace{-5pt}
We numerically study the effects of short-length PS on the SNR and on AIRs after fiber transmission by comparing different CCDM block lengths. The outer FEC is not included in the simulations as it does not affect the presented analysis. The simulated length of QAM symbols $x$ is 324000 symbols per polarization, corresponding to 30 or 20 FEC blocks (each of length 64800 bits) for 64QAM and 16QAM, respectively. Encoding is carried out with DVB-S2 LDPC codes. Each inner FEC block consists of several CCDM blocks whose length is varied from $n=10$ up to $n=5000$ amplitude symbols. The shaped amplitude distribution of 16QAM and 64QAM is fixed to $[0.7, 0.3]$ and $[0.4, 0.3, 0.2, 0.1]$, respectively. With this setup, all QAM transmit sequences have the same average distribution---the only difference is the length of the CCDM blocks that constitute that overall sequence. This approach allows to evaluate the effect of varying the DM block length on fiber NLI in an isolated manner.

For optical transmission, an idealized dual-polarization multi-span wavelength division multiplexing (WDM) fiber system is simulated. The shaped sequences in both polarizations are generated from independent random data, and each WDM channel is modulated individually. Two WDM setups with similar overall bandwidth are considered, comprising either 7 channels with 42 GBd symbol rate on a 50 GHz grid, or 5 channels at 64 GBd on a 75 GHz grid. The optimal per-channel transmit powers are 1~dBm and 2~dBm, respectively. When explicitly stated, a burst interleaver randomly permutes the information bits within each FEC block, and a symbol interleaver shuffles the QAM symbols. After root-raised cosine pulse shaping with 10\% rolloff, the dual-polarization WDM signal is transmitted over 10 spans of 80~km standard single-mode fiber ($\alpha=0.2$dB/km, $\gamma=1.37$~1/W/km, $D=17$~ps/nm/km). The entire span loss of 16~dB is compensated by an Erbium-doped fiber amplifier with 6 dB noise figure. Signal propagation over the fiber is simulated with the split-step Fourier method with 100~m step size. At the receiver, the center channel is ideally filtered using a matched filter, chromatic dispersion is compensated, and the nonlinearity-induced constant phase rotation is ideally compensated. Ten simulation runs are carried out for each setup and the considered figures of merit are averaged over all runs.

The effective SNR, averaged over both polarizations, is estimated from the unit-energy transmitted data $x$ and the received symbols $y$ as $1/\text{var}(y-x)$ where $\text{var}(\cdot)$ denotes variance. By considering the SNR only, the finite-length DM rate loss and the throughput improvement from PS are not considered. To take them into account, the AIRs for bit-metric decoding (BMD) and for a finite-length DM of length $n$ is computed as \cite[Appendix]{MPDM}
\begin{equation}\label{eq:air}
\setlength{\abovedisplayskip}{3pt}
\setlength{\belowdisplayskip}{3pt}
\textstyle
\AIRn = \left[ H(\mathbf{C}) - \sum_{i}^{m} H(C_i|Y) \right] - \Rlossn.
\end{equation}
The first part is the BMD rate with $\mathbf{C}=(C_1,\dots,C_m)$ representing the $m$ coded bit levels of the considered QAM format, $H(\cdot)$ denoting entropy and the channel output being $Y$. The rate loss \Rlossn is defined as the amplitude entropy minus the DM rate and in general decreases with $n$ \cite{GeorgTComm}. We note that \AIRn of \eqref{eq:air} is achievable with a finite-length DM that has rate loss \Rlossn and with capacity-achieving FEC. If an actual FEC is implemented, the required SNR for successful decoding will be decreased by approximately the FEC coding gap \cite{MPDM}.

\vspace{-5pt}
\subsection{Results}
\begin{figure}[t]
\begin{center}
\inputtikz{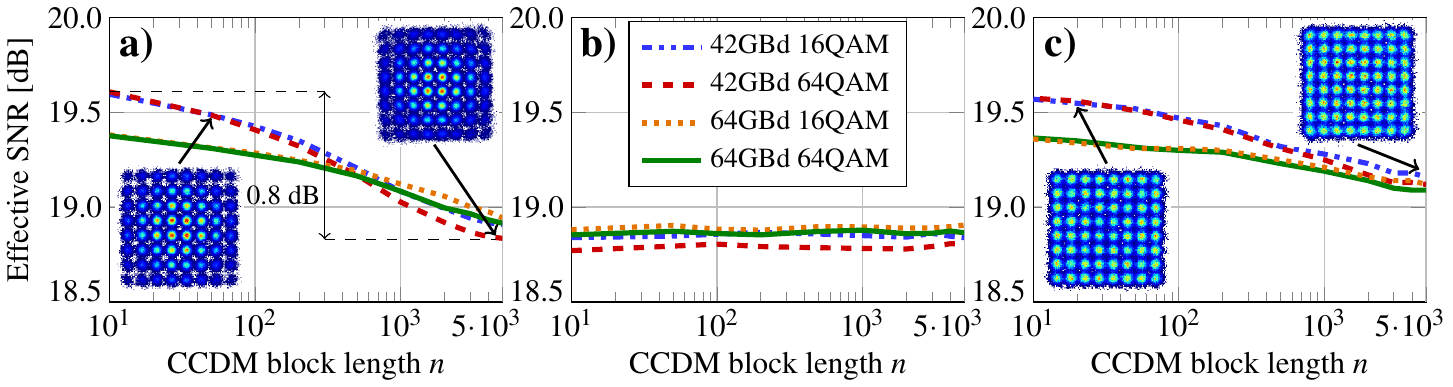}
\captionsetup{width=5.5in}
\vspace*{-0.9\baselineskip}
\caption{Effective SNR vs. CCDM length $n$. \textbf{a)} Shaped QAM with end-to-end (E2E) implementation without interleaving. \textbf{b)} E2E with interleaving \textbf{c)} E2E for uniform QAM generated by a CCDM.}
\label{fig:CCDM_SNR_vs_n}
\end{center}
\vspace*{-2.8\baselineskip}
\end{figure}

\vspace{-5pt}
In Fig.~\ref{fig:CCDM_SNR_vs_n}, the effective SNR in dB versus the CCDM block length $n$ in amplitude symbols is evaluated. Shown in Fig.~\ref{fig:CCDM_SNR_vs_n}~a) is the E2E implementation \emph{without} the interleavers. We observe that the effective SNR varies significantly with $n$. For 64QAM at 42~GBd, the SNR difference between $n=10$ and $n=5000$ is 0.8~dB, which is due to the mitigation of NLI at short $n$, as apparent from the constellations after DSP (insets). The reason for this effect is that certain sequences are impossible to occur when the overall transmit sequence consists of many short CCDM blocks. Recall that the constant-composition principle requires that every amplitude must occur a fixed number of times in a DM codeword. Thus, long runs of identical symbols are prohibited for short-length DM, or, equivalently, a temporal structure is inherently imposed in the output by short DMs, which is not guaranteed for long blocks. As an example, consider a CCDM with $n=10$ and the amplitude distribution $[0.4, 0.3, 0.2, 0.1]$ used for shaped 64QAM. The high-energy symbol (having probability 0.1) must occur exactly once every ten amplitudes. Also, there cannot be more than 8 consecutive occurrences of the low-energy symbol, which would be the case if one DM codeword ended with four occurrences of that amplitude and the next one began with this identical-amplitude run. 
In Fig.~\ref{fig:CCDM_SNR_vs_n}~b), the full E2E setup is simulated, now including the interleavers. We observe that the effective SNR is approximately constant over $n$ for all QAM formats and symbol rates because any temporal structure in the transmit sequence is broken up by the interleavers. When the entire PS system is emulated by directly generating the QAM symbols, no difference to E2E with interleaving is found (hence curves not depicted in Fig.~\ref{fig:CCDM_SNR_vs_n}~b).  
We note that any interleaving must be reversed before transmission in order to achieve temporal shaping gains. The operation of the burst interleaver can simply be undone after encoding and re-performed before decoding. The symbol interleaver, however, would have to be eliminated to preserve the NLI-mitigating properties, thereby losing diversity.
To highlight that the NLI mitigation is indeed the result of correlations in the transmit sequences, we simulated the E2E setup without interleaving for uniform QAM generated with a CCDM. Although practically not very relevant due to the rate loss, the block-length dependence of SNR for uniform QAM, shown in Fig.~\ref{fig:CCDM_SNR_vs_n}~c), demonstrates that it is indeed the temporal structure induced by short-length CCDM that leads to NLI mitigation. 

\begin{wrapfigure}{R}{0.4\textwidth}
\vspace*{-1.5\baselineskip}
\begin{center}
\inputtikz{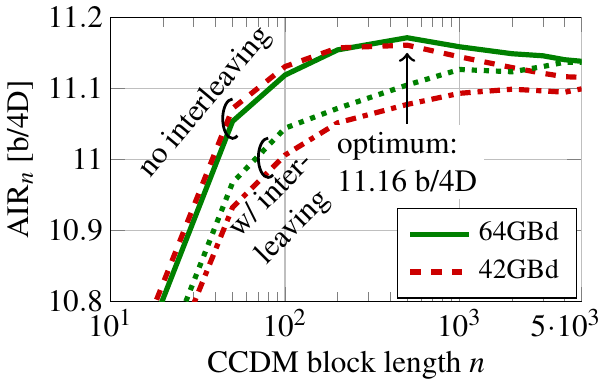}
\vspace*{-.5\baselineskip}
\caption{Achievable information rate \AIRn for shaped 64QAM vs. CCDM block length $n$.}
\label{fig:CCDM_GMI_vs_n}
\end{center}
\vspace*{-2\baselineskip}
\end{wrapfigure}

In Fig.~\ref{fig:CCDM_GMI_vs_n}, the finite-length-DM \AIRn in bits per 4D-symbol [b/4D] is shown versus $n$ for 64QAM and the E2E setups. When the interleavers are omitted, a maximum AIR of 11.16~b/4D is obtained at approximately 500 symbols where an optimal trade-off is achieved between short-length SNR improvement and rate loss reduction at large $n$. Note that the AIR optimum is more pronounced for advanced amplitude shapers that have a steeper rate loss decrease with $n$ \cite[Fig.~7]{ESS_Karim}. For blocks shorter than the optimum, the increased rate loss dominates over the SNR gains. At above-optimum $n$, the AIR is slightly reduced because of the smaller SNR gain, even though \Rloss is also decreased. Further shown in Fig.~\ref{fig:CCDM_GMI_vs_n} is the AIR with interleaving. Since the SNR is approximately constant (see Fig.~\ref{fig:CCDM_SNR_vs_n}~b), larger $n$ always gives higher AIRs.

\vspace{-9pt}
\section{Conclusions}
\vspace{-7pt}
We have shown in extensive fiber simulations that a short-length CCDM can give SNR improvements of up to 0.8~dB by introducing temporal structure in the transmit sequences that effectively mitigates fiber nonlinearities. This effect is present in end-to-end implementations of both shaped and uniform QAM with the DM included, yet cannot be observed when the PAS architecture is emulated only. The corresponding AIR improvements are rather small due to the large CCDM rate loss at short blocks. We have demonstrated that the NLI mitigation effect and related SNR gains can be destroyed if interleavers are employed and their operation is not reverted before transmission. Hence, CCDM-induced correlations must be preserved in order to achieve temporal shaping gains.\\[-1pt]
{\scriptsize
The work has been partially funded by the German Ministry of Education and Research in the projects SpeeD (\#13N1374) and PEARLS (\#13N14937).}


\end{document}